\documentclass[12pt]{iopart}
\usepackage[dvips]{graphicx}
\usepackage{iopams}

\usepackage[usenames]{color}

\newcommand{\old}[1]{}
\newcommand{\text}[1]{\mathrm{#1}}
\newcommand{\ii}{\mathrm{i}}

\begin{document}

\title{Rashba quantum wire: exact solution and ballistic transport}
\author{C A Perroni$^{1}$,
        D Bercioux$^{2,4}$, V Marigliano Ramaglia$^{3}$, and V~Cataudella$^{3}$}
\address{$^{1}$Institut f\"ur Festk\"orperforschung (IFF),
         Forschungszentrum J\"ulich,\\ D-52425 J\"ulich, Germany \\
         $^{2}$Institut f\"ur Theoretische Physik, Universit\"at
         Regensburg,\\ D-93040 Regensburg, Germany\\
         $^{3}$Coherentia-CNR-INFM and Dipartimento di Scienze
         Fisiche, Universit\` a degli Studi di Napoli ``Federico II'',
         I-80126 Napoli, Italy\\ $^{4}$ Physikalisches Institut,
         Albert-Ludwigs-Universit\" at, D-79104 Freiburg, Germany}

\date{\today}

\begin{abstract}
The effect of Rashba spin-orbit interaction in quantum wires with
hard-wall boundaries is discussed. The exact wave function and
eigenvalue equation are worked out pointing out the mixing between the
spin and spatial parts. The spectral properties are also studied
within the perturbation theory with respect to the strength of the
spin-orbit interaction and diagonalization procedure. A comparison is
done with the results of a simple model, the two-band model, that
takes account only of the first two sub-bands of the wire. Finally, the
transport properties within the ballistic regime are analytically
calculated for the two-band model and through a tight-binding Green
function for the entire system. Single and double interfaces
separating regions with different strengths of spin-orbit interaction
are analyzed injecting carriers into the first and the second
sub-band. It is shown that in the case of a single interface the spin
polarization in the Rashba region is different from zero, and in the
case of two interfaces the spin polarization shows oscillations due to
spin selective bound states.

\end{abstract}

\maketitle

\section{Introduction}
The \textit{Spintronics}~\cite{zutic:2004} is
one of the most prominent fields of  modern condensed matter
physics. Its target is to use spin to create electrical and
optoelectronic devices with new functionalities~\cite{wolf:2001}. Up
to now, starting form the seminal device by Datta and Das~\cite{Datta:1990}, 
several devices based on the giant and tunnel
magnetoresistance have been realized for read-head sensors and magnetic
random-access memories~\cite{wolf:2001}. The important task of the
integration of such spintronics technologies with the classical
semiconductor devices finds an obstacle in the small spin injection 
from magnetic to semiconductor materials
due to the large resistivity mismatch between magnetic and
semiconductor materials~\cite{schmidt:2000}. For this reason it is
useful to design semiconductor devices with an efficient all-electrical
spin-injection and detection via Ohmic contacts at the Fermi energy,
as it has been already realized for metallic
devices~\cite{jedema:2001,valenzuela:2006}.

Two important classes of spin-orbit interaction (SOI) are relevant for
semiconductor spintronics: the Dresselhaus
type~\cite{dresselhaus:1955} and the Rashba type~\cite{rashba:1960} coupling.
The former arises from the lack of symmetry in the bulk inversion
whereas the latter arises from the asymmetry along the
growing-direction-axis of the confining quantum well electric
potential that creates a two-dimensional electron gas (2DEG) on a
narrow-gap semiconductor surface. Since the Rashba SOI can be tuned by
an external gate
electrode~\cite{nitta:1997,schaepers:1998,grundler:2000} it is
envisaged as a tool to control the precession of the electron spin in
the Datta-Das proposal for a field-effect spin
transistor~\cite{Datta:1990}.

Quasi-one-dimensional electron gases or quantum wires (QWs) are
realized by applying split gates on top of a 2DEG in a semiconductor
heterostructure~\cite{schaepers:2004}. The main effect owing to the
confining potential is quantization of the electron motion in the
direction orthogonal to the wire axis. The combination of this
confining potential and the Rashba SOI gives rise to sub-band
hybridization that can affect the working principle of the
field-effect spin transistor. Mireles and
Kirczenow~\cite{Mireles:2001} have numerically studied this effect and
they have shown that a large value of the Rashba SOI can produce
dramatic changes in the transport properties of the device till to
suppress the expected spin modulation. The effect of sub-band
hybridization has been investigated by Governale and Z\"
uelike~\cite{governale:2002} in a QW with parabolic confinement. They
show that electrons with large wave vectors in the lowest spin-spit
sub-bands have essentially parallel spin. But in proximity of the
anti-crossing points due to the sub-band hybridization it is no more
appropriate to use the spin quantum number in order to characterize
the electron state in the QW. Furthermore, they show that it is not
possible to transfer the finite spin polarization of the QW to some
external leads.

In this Article we study the spectral and the transport properties of
a QW in the presence of SOI. The main result of this Article is to show
how to achieve non-zero spin polarization in external leads using spin
unpolarized injected carriers. This can be obtained by injecting carriers
in all the active sub-bands due to the quantum confinement. At the
opening of each new sub-band the hybridization owing to SOI gives rise
to spin selective bound states reflecting in a oscillating spin
polarization. Here, we want to stress that those spin polarized bound
states are not in contradiction with any fundamental symmetry property
of the system~\cite{zhai:2005}.

This Article is organized in the following way: in Sec.~\ref{sec:one}
we evaluate the spectral properties using the wave function
approach~\cite{Marigliano1,marigliano:2004,bercioux:2004}. In Sec.~\ref{sec:two} 
we provide an exact calculation for the
spectral properties  investigated within the perturbation theory
approach and with the exact diagonalization in a truncated Hilbert
space. Here we also introduce a minimal model featuring the basic
characteristic of a QW with SOI named
\emph{two-band} model. This is used in Sec.~\ref{sec:three} in order
to study the transport properties of a QW in presence of a
single interface between a region with and without SOI and in the case of
a double interface (spin-field effect transistor scheme).  Conclusions
are ending the Article.

\section{Exact solution of Rashba quantum wire: wave-function~\label{sec:one}}

Let us consider a 2DEG filling the plane ($x,z$). The charge carriers
have momentum $\vec{p} \equiv (p_x,p_z)$ and effective mass $m$. The
particles are confined along the $z$-direction by the potential $V(z)$
and subjected to the Rashba spin-orbit interaction (SOI). The
single-particle Hamiltonian reads
%
%
\begin{equation}
\mathcal{H}=\frac{1}{2m}\left( p_x^2+p_z^2\right) +V(z)+ \mathcal{H}_\text{R},  
\label{eq1}
\end{equation}
%
%
where $\mathcal{H}_\text{R}$ is the Rashba SOI
%
%
\begin{equation}
\mathcal{H}_\text{R}=\frac{\hbar k_\text{SO}}{m} \left( \sigma _z p_x
- \sigma _xp_z\right). 
\label{eq2}
\end{equation}
%
%
In Eq.(\ref{eq2}) $\sigma_x$ and $\sigma_z$ are the $x$ and $z$
components, respectively, of the vector $\vec{\sigma}$ of Pauli
matrices, and $k_\text{SO}$ is the SOI constant. This can be tuned by
means of external gates perpendicular to the 2DEG
\cite{nitta:1997,schaepers:1998,grundler:2000}.

In the following we assume that the potential $V(z)$ provides a
confinement with hard walls at $z=0$ and $z=W$.  The strategy to find
the wave-function of the Hamiltonian (\ref{eq1}) is similar to the
procedure followed in the absence of SOI: one exactly solves the 2D
problem, then considers the quantizing effect of confinement on the
wave function $\psi(x,z)$. In the first subsection we shortly recall
the spin-dependent solution of the 2DEG with SOI, then we impose the
boundary conditions $\psi(x,z=0)=\psi(x,z=W)=0$. In the presence of
SOI, these relations mix the $z$ part of the wave-function with its
spinor component.

\subsection{Solution without confinement}
Without confinement both components of the momentum $\vec{p}=\hbar
\vec{k} $ are conserved. The eigenfunctions of the Hamiltonian
(\ref{eq1}) in the absence of confining potential $V(z)=0$ are denoted
by the two spin modes $(+)$ and $(-)$ and read
%
%
\numparts
\begin{eqnarray}
\psi _{\vec{k},+}\left( x,z\right) & = &\exp \left[
\ii(k_xx+k_zz)\right] \left( 
\begin{array}{c}
\cos(\theta/2) \\ 
-\sin(\theta/2)
\end{array}
\right), 
\label{eq3a}\\
\psi _{\vec{k},-}\left( x,z\right)  & = &\exp \left[
\ii(k_xx+k_zz)\right] \left( 
\begin{array}{c}
\sin(\theta/2) \\ 
\cos(\theta/2)
\end{array}
\right),  
\label{eq3b}
\end{eqnarray}
\endnumparts
%
%
whose corresponding eigenvalues are given by 
%
%
\begin{equation}
E_{\pm }=\frac{\hbar ^2}{2m}\left( k^2 \pm 2k k_\text{SO}  \right),  
\label{eq4}
\end{equation}
%
%
where $k=\sqrt{k_x^2+k_z^2}$ is the modulus of the wave-vector in
$x$-$z$ plane, $k_x=k \cos(\theta)$, $k_z=k \sin(\theta)$, with
$\theta$ the angle formed between the vector $\vec{k}$ and the $x$
axis. It is clear for Eqs. (\ref{eq3a},\ref{eq3b}) that the spinors
$\chi _{\pm }$ of the two modes are orthogonal to each other.  We
remind that the Rashba SOI can be viewed as a magnetic field parallel
to the ($x,z$) plane and orthogonal to the wave-vector $\vec{k}$. The
net effect is to orientate the spin along the direction perpendicular
to the wave-vector \cite{Marigliano1}.

%
%
\begin{figure}
        \begin{minipage}[l]{0.45\textwidth} \centering
        \includegraphics[scale=0.7,angle=0]{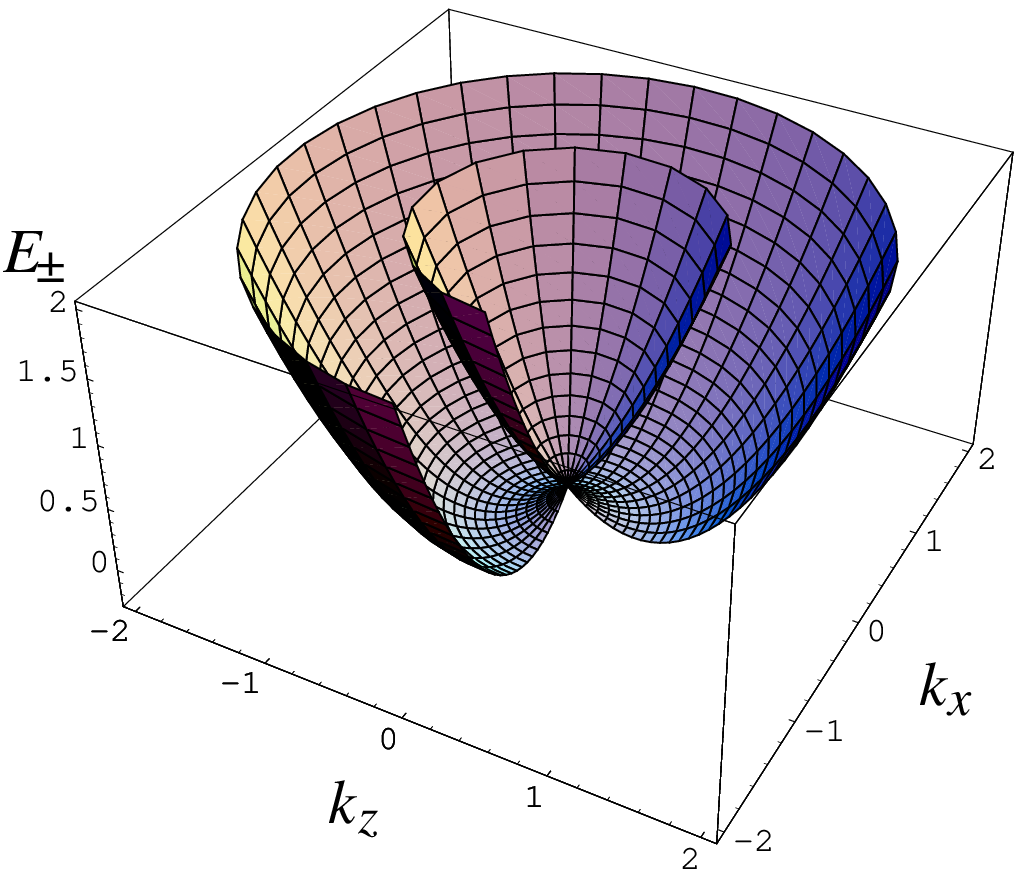}
        \end{minipage} \begin{minipage}[r]{0.45\textwidth} \centering
        \includegraphics[scale=0.45]{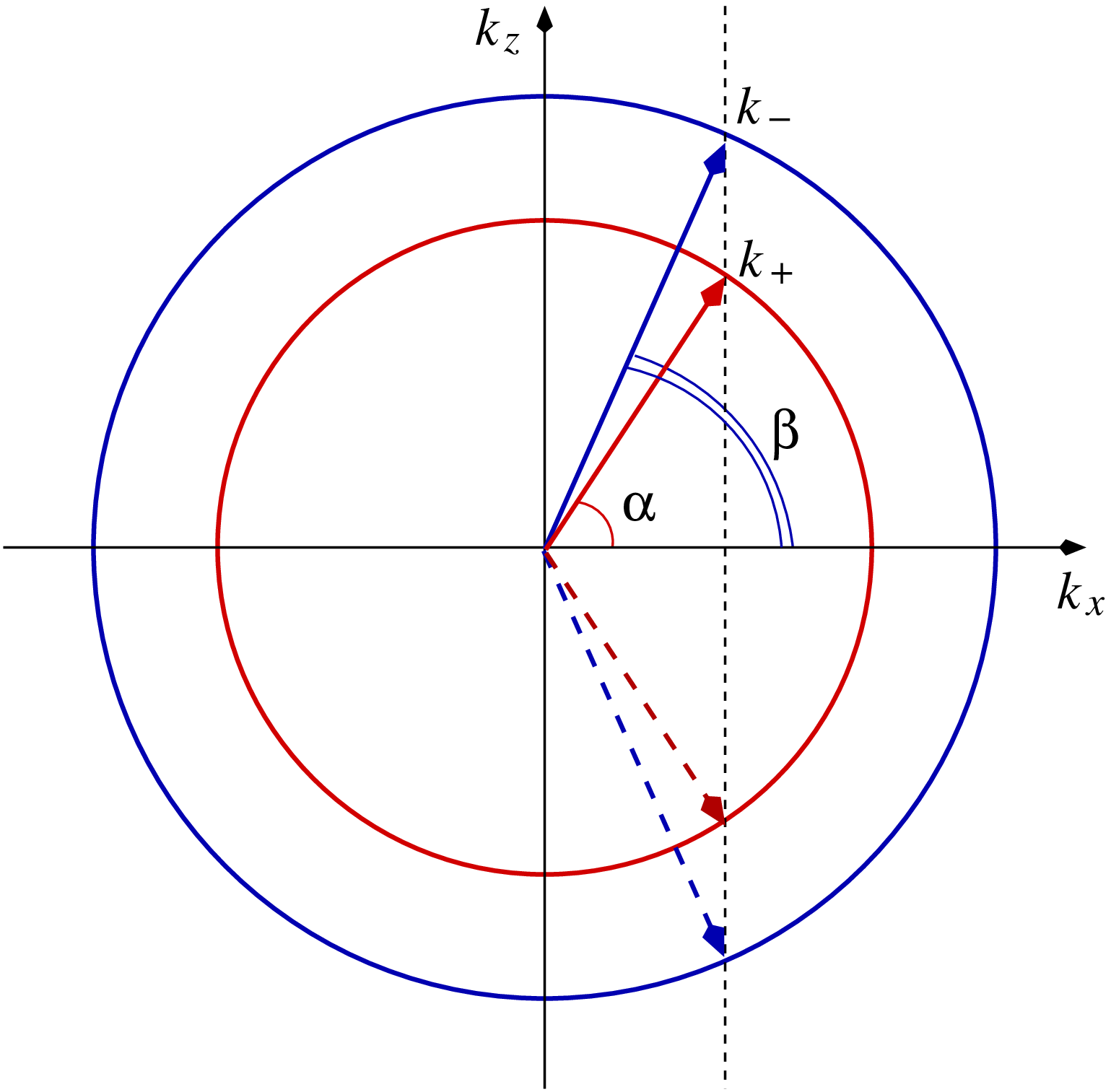} \end{minipage}
        \caption{(Left Panel) Spectrum of the two-dimensional electron gas in
        the presence of SOI interaction as a function of the
        wave-vectors $k_x$ and $k_y$. At fixed positive energy, the
        wave-vectors of the mode $(+)$ and $(-)$ are on two concentric
        circles (radius for $(-)$ mode larger than that for the $(+)$
        mode).(Right Panel) At fixed positive energy and x component of the
        wave-vector, the 4 possible values of the z component are
        shown.}  \label{fig_1_d}
\end{figure}
%
%
In order to determine the solution with confinement, it is essential
to find the eigenfunctions in the free case when the total energy $E$
and the momentum along $k_x$ are fixed (See Fig.~\ref{fig_1_d}).
Fixing the total energy $E$ , we note that there are two values of
total momentum $k$, corresponding to the different modes, fulfilling
the Eq.~(\ref{eq4}) expressed as a linear combination of those four
waves:
%
%
\begin{equation}
k_{\pm }=\sqrt{\frac{2m}{\hbar^2}E+k_\text{SO}^2}\mp k_\text{SO}.
\label{eq5}
\end{equation}
%
%
The propagation directions for the $k_\pm$ modes are fixed by the
momentum $k_x$. For $E>0$, the mode ($+$) is characterized by
the propagation direction $\pm\alpha = \arccos(k_x/k_+)$ fixing the
value of $k_z=\pm k_+\sin(\alpha)$, whereas the mode ($-$) has
propagation direction $\pm\beta = \arccos(k_x/p_-)$ and $k_z=\pm
k_-\sin(\beta)$.  Therefore, the generic wave function
$\psi_{E,k_x}(x,z)$ is given by
%
%
\begin{equation}
\psi _{E,k_x}\left(x,z\right) =\e^{\ii k_x x} \left[ 
A \psi_{1,E}^{(+)}(z)+B \psi_{2,E}^{(+)}(z)+
C \psi_{1,E}^{(-)}(z)+D \psi_{2,E}^{(-)}(z)
\right],
\label{eq3az}
\end{equation}
with
\begin{equation}
\psi_{\ell,E}^{(+)}(z) =  \e^{- \ii (-1)^{\ell} k_+ z \sin(\alpha)}
\left( 
\begin{array}{c}
\cos(\alpha/2) \\ 
 (-1)^\ell \sin(\alpha/2)
\end{array}
\right),
\label{eq3avzz}
\end{equation}
\begin{equation}
\psi_{\ell,E}^{(-)}(z) =\e^{-\ii (-1)^{\ell} k_- z \sin(\beta)}
\left( 
\begin{array}{c}
- (-1)^{\ell} \sin(\beta/2) \\ 
\cos(\beta/2)
\end{array}
\right),
\label{eq3axww}
\end{equation}
%
%
and $\ell=1,2$.

For $-k_- \leq k_x < -k_+$ or $k_+ < k_x \leq k_-$, the wave-function
(\ref{eq3az}) is still valid. However, one has $\alpha=\ii a$,
implying that $\cos(\alpha)=\cosh(a)$ and $\sin(\alpha)=\ii\sinh(a)$,
therefore the ($+$) mode becomes an evanescent one. Moreover, if $k_x
> k_-$ or $k_x < -k_-$, then $\beta=\ii b$ and also the ($-$) mode
changes into an evanescent one.

For $E<0$ the 4 values of $k_z$ are only relative to modes $(-)$. A
wave-function similar to (\ref{eq3az}) can be written. Also in
this case the Fermi surface is formed by two circles but now they
correspond to the same energy $E_-$.  However, in the next section, we
will see that, from weak to intermediate values of the Rashba SO
coupling, only positive values of the energy are important due to the
effect of the confinement.

\subsection{Solution with confinement}
The wave-function (\ref{eq3az}) represents the starting point for
taking account of the confinement. In fact, the hard wall boundary conditions
are obtained by imposing that the wave-function is zero on the borders
($z=0$ and $z=W$): $\psi_{E,k_x}(x,z=0)=\psi_{E,k_x}(x,z=W)=0$. For
$E>0$, we get the following exact eigenvalue equation for the Rashba
quantum wire
%
%
\begin{eqnarray}
1-\cos[k_+W\sin(\alpha)] \cos[k_-W\sin(\beta)]+
\nonumber \\
\sin[k_+W\sin(\alpha)] \sin[k_-W\sin(\beta)] 
\frac{[1+\cos(\alpha)\cos(\beta)]}{\sin(\alpha)\sin(\beta)}=0.
\label{eq3at}
\end{eqnarray}
%
%
Therefore, via the SOI, the quantities $k_+$ and $k_-$, and 
clearly the energy, are related to the spinor components of the 
wave-function. A similar equation is valid for $E<0$. 

In the absence of SOI, the quantized energy levels are independent of
spin behavior. Actually, one gets $k_{\pm}=\sqrt{2mE/\hbar^2}$ and
$\alpha=\beta$. This yields
%
%
\begin{equation}
\sqrt{\frac{2mE}{\hbar^2}}\, W \sin(\alpha)=n \pi, 
\end{equation}
%
%
with $n$ being a positive integer number, so that, together with the relation
$\cos(\alpha)=k_x/k_\pm$, we obtain
%
%
\begin{equation}
E_n=\frac{\hbar^2}{2m}\left(\frac{n^2 \pi^2}{W^2}+k_x^2\right),
\end{equation}
%
%
which are the energy values for the sub-bands of the quantum wire without SOI.

In the presence of SOI, an important limit is obtained for 
$k_x=0$. Indeed the eigenvalue equation becomes
%
%
\begin{equation}
        \cos[(k_+ + k_-)W]=1,
\end{equation}
%
%
implying  that 
%
%
\begin{equation}
E_n=\frac{\hbar^2}{2m}\left( \frac{n^2 \pi^2}{W^2}- k_\text{SO}^2 \right).
\label{eqk0}
\end{equation}
%
%
Therefore, all the sub-bands are shifted down by the SOI term
$k_\text{SO}^2$.  For values of the SOI such that $k_\text{SO}<\pi/W$,
the energies are positives and the wave function (\ref{eq3az}) holds
true. This means that the spin-precession length $L_\text{SO}=\pi
/k_\text{SO}$ has to be larger than the wire width.

\section{Two-band model and perturbation theory~\label{sec:two}}

In the previous section we started from the wave-function of the 2D
model with SOI, and then imposed the conditions due to
confinement. Now, we consider the opposite point of view. First we
take into account the exact solution of the quantum wire in the
absence of the SOI, then we study its effect on the
sub-bands. Because of SOI, a coupling between sub-bands with
opposite spins occurs. In order to study the effects of this coupling,
in the first subsection we will discuss the results within the first-
and second-order perturbation theory approach with respect to the
SOI. In the second subsection we consider the two-band model, where
only the first two bands of the unperturbed spectrum are assumed to be
coupled by the interaction. This assumption is valid if the
wire is very narrow. Moreover, this simple system is studied since it
provides a simple understanding of the transport properties. Finally,
in the third subsection, we will show the results of the exact
diagonalization of the model.
    
The Hamiltonian (\ref{eq1}) is considered to be split into two terms:
$\mathcal{H}_0$ and $\mathcal{H}_\text{R}$. The term $\mathcal{H}_0$
is simply the Hamiltonian of the wire without SOI:
%
%
\begin{equation}
\mathcal{H}_0=\frac{1}{2m}\left( p_x^2+p_z^2\right) +V(z),  
\label{eq1bis}
\end{equation}
%
%
where $V(z)$ is the hard-wall confining potential. Due to the presence
of the potential $V(z)$, only the momentum $p_x=\hbar k_x$ is
conserved. We find the matrix elements of $\mathcal{H}$ in the basis
of $\mathcal{H}_0$ indicated by $|k_x,n,\sigma\rangle$, with $n$ index
of the sub-band and $\sigma=\pm1$ for up or down spin,
respectively. The SOI term contains terms $\sigma_z p_x $ and
$-\sigma_x p_z$.  For $\ell=n$, only the former term of
$\mathcal{H}_\text{R}$ is acting on the unperturbed states with the
same spin state, so that the matrix elements are
%
%
\begin{equation}
\langle k_x,\ell,\sigma | \mathcal{H}_\text{R} | k_x,n,\sigma' \rangle =
\frac{\hbar^2 k_\text{SO} k_x}{m} \, \sigma' \delta_{\sigma,\sigma'},
\end{equation}
%
%
while, for $\ell \neq n$, the latter term $\mathcal{H}_\text{R}$ 
couples sub-bands with opposite spin and parity, so that the
matrix elements are
%
%
\begin{equation}
\langle k_x,\ell,\sigma | \mathcal{H}_\text{R} | k_x,n,\sigma' \rangle
=J_{\ell,n} \delta_{\sigma,- \sigma'},
\end{equation}
%
%
with $J_{\ell,n}$ independent of the wave-vector $k_x$
%
%
\begin{equation}
J_{\ell,n}=\frac{\ii \hbar^2 k_\text{SO}}{m\, W} \frac{2 \ell n}{\ell^2-n^2}
\left[ 1-(-1)^{|\ell-n|} \right].
\end{equation}
%
%
If we express the energies in the unit $\hbar^2/2mW^2$, the lengths in
$W$ and the wave-vectors in $1/W$, we recast the following matrix
elements for the entire Hamiltonian $\mathcal{H}$:
%
%
\begin{eqnarray}
\langle k_x,\ell,\sigma| \mathcal{H} |k_x,n,\sigma' \rangle &=&
\left[{\bar E}_n^{(0)}({\bar k}_x)+2  
{\bar k_\text{SO}} {\bar k_x} \sigma'\right] \delta_{\ell,n}
\delta_{\sigma,\sigma'}+
\nonumber \\
&& 
{\bar J}_{\ell,n}[1-\delta_{\ell,n}] \delta_{\sigma,-\sigma'},
\label{eqadi}
\end{eqnarray}
%
%
where ${\bar k}_x=k_x W$, 
${\bar E}_n^{(0)}({\bar k}_x)= {\bar k}_x^2+n^2 \pi^2$, 
${\bar k}_\text{SO}=k_\text{SO} W$, and ${\bar J}_{\ell,n}$ 
proportional to the dimensionless SOI term  ${\bar k}_\text{SO}$  
%
%
\begin{equation}
{\bar J}_{\ell,n}=\ii {\bar k}_\text{SO} \frac{4 \ell n}{\ell^2-n^2}
\left[ 1-(-1)^{|\ell-n|} \right].
\end{equation} 
%
%

\subsection{Perturbation theory}
The correction to the unperturbed energies $\bar{E}_n^{(0)}$ within
the first-order perturbation theory is simply derived considering only
the diagonal terms of Eq.(\ref{eqadi}). Therefore, at first-order, the
$n$-th sub-band is simply affected by the spin slitting due to the
contribution $\sigma_z p_x$ of $\mathcal{H}_\text{SO}$:
%
%
\begin{equation}
{\bar E}_{n,\sigma}^{(1)}({\bar k}_x)={\bar E}_n^{(0)}({\bar k}_x)+2 
{\bar k_\text{SO}} {\bar k_x} \sigma,
\end{equation}
%
%
and eigenvectors equal to those of the unperturbed system. This
splitting controlled by SOI gives rise to a first-order spectrum with
crossings between sub-bands with opposite spins. For example, the
first and second sub-band intersect at ${\bar k}_x= \pm 3 \pi^2/4
\bar{k}_\text{SO}$ and the others for larger values of ${\bar
k}_x$. This suggests that the full effect of the interaction should
remove this crossing by mixing the behavior of coupled sub-bands. Due
to the presence of those level crossings, the correction to the energy
levels within the second-order perturbation theory fails for values of
${\bar k}_x$ close to intersections. Far from the crossing points, it
is easy to derive the contribution in the second-order to the energy
%
%
\begin{equation}
{\bar E}_{n,\sigma}^{(2)}=\sum_{\ell(\neq n)}\sum_{\sigma'(\neq \sigma)}
\frac{ |{\bar J}_{\ell,n}|^2}{\pi^2 (n^2-\ell^2)}=-{\bar k}_\text{SO}^2,
\end{equation}
%
%
a quantity independent of $k_x$, $n$ and $\sigma$. This result is
indubitably valid for ${\bar k}_x=0$. Indeed, it coincides with the
result (\ref{eqk0}) obtained in the previous section by using the
exact wave-function. This shows that at ${\bar k}_x=0$ the energy
correction within the second-order perturbation theory is able to
fully describe the energy spectrum. Also, the correction of the
wave-function at first-order can be evaluated. If at zero-order the
spin is $\sigma$, at first-order one takes contribution from
$-\sigma$:
%
%
\begin{equation}
\psi_{n,\sigma}^{(1)}(z)=\frac{{\bar k}_\text{SO}}{4 \pi^2}\sqrt{\frac{2}{W} }
 \left[ S_{1,n} (z) -S_{2,n} (z) \right]
|-\sigma \rangle\,,
\end{equation}  
%
%
with
%
%
\begin{equation}
S_{1,n} (z)=
\Phi \left( e^{-2 i \pi z /L}, 2, \frac{-n}{2} \right)-
\Phi \left( e^{-2 i \pi z /L}, 2, \frac{n}{2} \right),
\end{equation}
%
%
and
%
%
\begin{equation}
S_{2,n} (z)=
\Phi \left( e^{2 i \pi z /L}, 2, \frac{-n}{2} \right)-
\Phi \left( e^{2 i \pi z /L}, 2, \frac{n}{2} \right),
\end{equation}  
%
%
where $\Phi(x,s,a)=\sum_{k=0}^{\infty} x^k/(a+k)^s$ is the Lerch transcendent 
function~\cite{Laurincikas:2003}.

\subsection{Two-band model}
In order to investigate the effects of the coupling between sub-bands 
induced by SOI, it is convenient to analyze the two-band model that 
 will be also considered in the section devoted to transport 
properties. This model takes only the first and the second sub-band of the unperturbed
wire into account . The $4 \times 4$ problem can be decoupled into two $2 \times 2$
problems. The only thing to evaluate is ${\bar J}_{1,2}=-\ii 16 {\bar
k}_\text{SO}/3$.  One gets $4$ eigenvalues~\cite{Mireles:2001,governale:2002}:
%
%
\begin{equation}
\epsilon_{1+}({\bar k}_x)=\frac{5 \pi^2}{2}+{\bar k}_x^2-g_1({\bar k}_x), 
\,\,\epsilon_{1-}({\bar k}_x)=\epsilon_{1+}({-\bar k}_x),
\end{equation} 
%
%
\begin{equation}
\epsilon_{2+}({\bar k}_x)=\frac{5 \pi^2}{2}+{\bar k}_x^2-g_2({\bar k}_x), 
\,\,\epsilon_{2-}({\bar k}_x)=\epsilon_{2+}({-\bar k}_x),
\end{equation}
%
%
with 
%
%
\begin{equation}
g_1({\bar k}_x)=\frac{1}{2} \sqrt{ (3 \pi^2 - 4 {\bar k}_x {\bar k}_\text{SO})^2
+\frac{1024 {\bar k}_\text{SO}^2}{9}},\,\,\, 
g_2({\bar k}_x)=g_1({-\bar k}_x).
\end{equation} 
%
%
The eigenvectors can also be calculated. For example, the eigenvector 
corresponding to $\epsilon_{1+}$ is
%
%
\begin{equation}
\psi_{1+}(x,z)=e^{i k_x x} \sqrt{\frac{2}{W}} 
\frac{1}{\sqrt{1+\left[ f_1\left( {\bar k}_x \right) \right]^2}}
\left( 
\begin{array}{c}
\sin \left(\frac{\pi z}{W}\right) \\ 
\ii f_1 \left( {\bar k}_x \right) \sin\left( \frac{2 \pi z}{W} \right)
\end{array}
\right),
\label{eqe1+}
\end{equation}
%
%
with $f_1 \left( {\bar k}_x \right)$ given by
%
%
\begin{equation}\label{fun:f:one}
f_1 \left( {\bar k}_x \right)=\frac{3}{16 {\bar k_\text{SO}}} 
\left[ -\frac{3 \pi^2}{2} + 2 {\bar k}_\text{SO} {\bar k}_x +g_1({\bar k}_x)
\right].
\end{equation}
%
%

%
%
\begin{figure}
        \begin{center} 
        \includegraphics[scale=0.3]{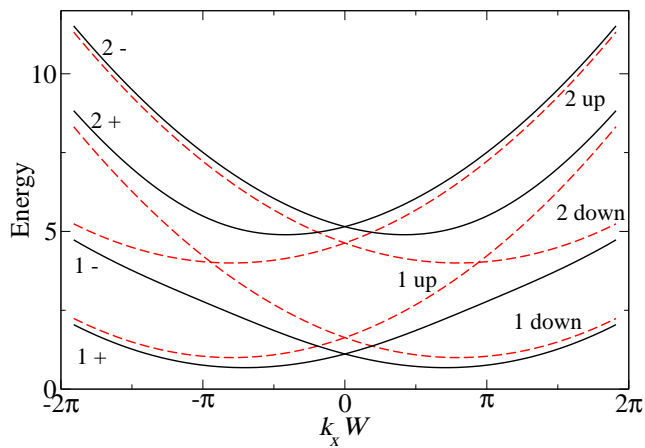}
        \caption{Energy levels of the two-band model (in units of
        $\hbar^2/2mW^2$) as a function of the dimensionless wave-vector
        $k_x W$ for the value $k_\text{SO}W = \pi$ of the dimensionless
        spin-orbit parameter. The spectrum derived from the
        diagonalization of the two-band system, indicated in figure by $1+$, 
        $1-$, $2+$, and $2-$ (solid lines), is compared with that 
        from first-order perturbation theory, indicated by 
        $1$ up, $1$ down, $2$ up, $2$ down (dashed lines).}  \label{fig_1_a} \end{center}
\end{figure}
%
%
As shown in Fig.~\ref{fig_1_a}, the eigenvalues (solid lines) do not
show any intersection for $k_x$ different from zero. Therefore, the
inter-band coupling removes the crossings of the first-order
perturbation theory solution (dashed line). As a result, the energy
eigenstates are no longer eigenstates of
$\sigma_z$~\cite{governale:2002} and the spin state depends on the
wave-vector $k_x$. Close to the crossing point, the wave function of
$1 \uparrow$ and $2 \downarrow$, for example, are strongly mixed in
the mode $1+$. However, far from the intersection, the mode given by
the diagonalization preserves the original behavior of the component
wave-functions. For example, if we analyze the behavior of the
eigenstate $\psi_{1+}(x,z)$, we get
%
%
\begin{equation}
\lim_{k_x \to -\infty}\psi_{1+}(x,z)=\psi_{1 \uparrow}(x,z),\,\,
\lim_{k_x \to \infty}\psi_{1+}(x,z)=\psi_{2 \downarrow}(x,z).
\end{equation}
%

The behavior of the two-band model shows a general trend: only taking
into account the coupling between sub-bands the description is qualitatively correct. 
The crossing are artifacts of the lowest-order perturbation
theory. The spectrum within the two-band model is reliable only for
very narrow wires. In the general case, the low-energy description
given by this model is too poor for the bands $2+$ and $2-$. This can
be easily seen if one considers the energy values at $k_x=0$. In fact one gets
for sub-bands $1 \pm$ and $2 \pm$, respectively,  the achieved values are
%
%
\numparts
\begin{eqnarray}
\epsilon_{1 \pm}=\frac{5 \pi^2}{2}- \frac{1}{2} 
\sqrt{9 \pi^4+ \frac{1024 {\bar k}_\text{SO}^2}{9} }\,, \\
\epsilon_{2 \pm}=\frac{5 \pi^2}{2}+\frac{1}{2} 
\sqrt{9 \pi^4+ \frac{1024 {\bar k}_\text{SO}^2}{9} }.
\end{eqnarray}
\endnumparts
%
%
In the limit of small ${\bar k}_\text{SO}$, they become
%
%
\numparts
\begin{eqnarray}
\epsilon_{1 \pm}=\pi^2- \frac{256}{27 \pi^2} 
{\bar k}_\text{SO}^2 \simeq \pi^2 -0.961{\bar k}_\text{SO}^2 \,,\\
\epsilon_{2 \pm}=4 \pi^2+\frac{256}{27 \pi^2} 
{\bar k}_\text{SO}^2 \simeq 4\pi^2 +0.961{\bar k}_\text{SO}^2.
\end{eqnarray}
\endnumparts
%
%
From the comparison with the exact solution we find
that the lowest sub-bands acquire a correction with
the right sign and very close to the
exact result, while the upper sub-bands have even the wrong
sign. Therefore, in order to give a reasonable description of the
low-energy part of the spectrum, more bands are necessary. 
This will also play an important role in the transport properties. 

%
%
\begin{figure}
        \begin{center} 
        \includegraphics[scale=0.3]{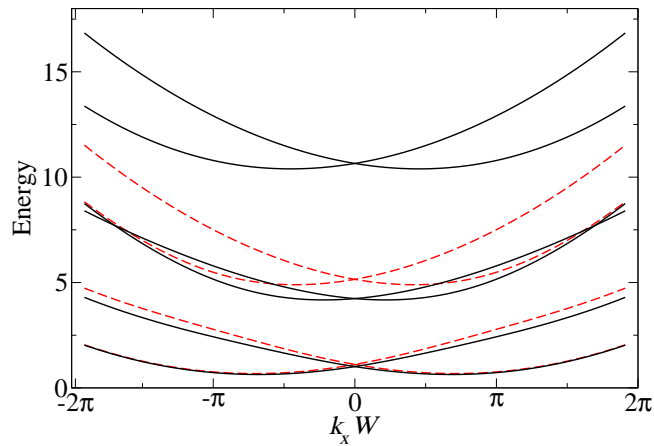}
        \caption{Energy levels of the wire (in units of
        $\hbar^2/2mW^2$) as a function of the dimensionless wave-vector
        $k_x W$ for the value $k_\text{SO}W = \pi$ of the dimensionless
        spin-orbit parameter. The spectra derived from the
        diagonalization of the system with 3 sub-bands (solid line)
        and 2 sub-bands (dashed line) are shown.}  \label{fig_2_a}
        \end{center}
\end{figure}
%
%
\subsection{Exact diagonalization}
In order to verify the importance of including more than two
sub-bands, one can directly diagonalize the Hamiltonian of the
system~\cite{governale:2002}. This can be done considering the matrix
elements (\ref{eqadi}). In Fig.~\ref{fig_2_a} we consider the
diagonalization in the subspace of 3 spin degenerate bands. It is
apparent that already at this level the corrections to the level
energies and wave-functions are important for the second sub-band. For
example, it is very close to the correct behavior at
$k_x=0$. Considering $l$ sub-bands for the diagonalization, one is
able to get a reliable behavior starting from sub-band 1 to $l-1$.

\section{Ballistic Transport\label{sec:three}}
In this section the issue is to study the quantum transport
properties within the ballistic regime. The standard
Landauer-B\"uttiker formalism will be employed. We will start
considering a wire divided in two different regions: one with SOI and
one without. Then we take the case of a quantum wire with a finite SOI
region into account. The results are obtained within the
approximation of the two-band model and those results will be
compared with a numerical tight-binding method.

\subsection{Single interface}
We consider a QW divided in two main regions: in the right
region ($x>0$) SOI is present while in the left region ($x<0$) it is
not. The interface separating the two regions is considered to be sharp and
is described by a $\delta$-like potential. The Hamiltonian of this
hybrid system reads
%
%
\begin{eqnarray}
\mathcal{H}_\text{hyb}&=&{\vec p}\frac{1}{2m(x)}{\vec p} +V(z)+\frac
{\hbar k_\text{SO}(x)} {m_\text{SO}}  
\left( \sigma _z p_x-\sigma _x p_z\right) \nonumber \\
&&-i\sigma_z \frac{\hbar}{2m_\text{SO}}\frac{\partial
k_\text{SO}(x)}{\partial x}+\frac{\hbar^2u}{2m(x)}\delta(x).   
\label{eqhyb}
\end{eqnarray}

%
%
We assume that the mass and the strength of the SOI are piecewise
constant with $k_\text{SO}(x)=k_\text{SO} \theta(x)$. For simplicity
the mass is considered equal on both sides of the interface. The
fourth term is necessary to get $\mathcal{H}_\text{hyb}$ hermitian. At
the interface the spinor eigenstates of $\mathcal{H}_\text{hyb}$ have
to be continuous, whereas their derivatives have a discontinuity $
u-i\sigma_z k_\text{SO}$ due to the SOI and to the $\delta$-like
potential in $x=0$:
%
%
\numparts
\begin{eqnarray}
\psi(0^+)=\psi(0^-)\label{eqhyb2a}\,, \\ 
\left.\frac{\partial \psi(x)}{\partial x}\right|_{x=0^+} -
\left.\frac{\partial \psi(x)}{\partial x}\right|_{x=0^-}= (u-\ii\sigma_z
k_\text{SO}) \psi(0).\label{eqhyb2b}
\end{eqnarray}
\endnumparts
%
%

In order to study the effects of the sub-band hybridization on the
transport properties, we start considering the injection of carriers
only within the first sub-band. This can be achieved requiring that
the second sub-band is behaving as an evanescent wave. In this context
the Eqs. (\ref{eqhyb2a}-\ref{eqhyb2b}) are reduced to a set of two
decoupled systems of four times four equations for the variables
$r_{1+}$, $t_{1+}$, $r_{2-}$, $t_{2-}$ and $r_{1-}$, $t_{1-}$,
$r_{2+}$, $t_{2+}$ respectively, where $t_{1(2),+(-)}$ and
$r_{1(2),+(-)}$ are the transmission and the reflection amplitudes in
the first (second) sub-band with spinor $+$ ($-$) respectively.  The
knowledge of the transmission and reflection amplitudes permits to
evaluate the probability current and, as a consequence, the
transmission probabilities for spin-up and spin-down carriers. Due to
the presence of SOI only for $x>0$, the probability current
has two different forms given by
%
%
\begin{equation}
\vec{j} = \displaystyle\frac{1}{m} \left\{
\begin{array}{l r}
\Re \left\{ \psi^\dag\vec{p}\, \psi \right\} & \text{for}~x<0\\
\Re \left\{\psi^\dag\left[\vec{p} + \hbar k_\text{SO} \left( \hat{y}
\times \vec{\sigma} \right) \right] 
 \psi \right\} & \text{for}~x>0
\end{array}
\right.
\end{equation}
%
%
where $\psi$ is the wave function solution of the system of
Eqs.~(\ref{eqhyb2a}-\ref{eqhyb2b}). A direct evaluation of the
transmission probabilities~\cite{ferry:1997} results in the following
expressions:
%
%
\numparts
\begin{eqnarray}
T_\uparrow = \frac{|t_{1+}|^2}{k_\text{in}} \left[ k_{1+}(k_\text{in}) +
k_\text{SO} \langle \sigma_z \rangle_{1+} \right] \,,\label{prob:up}\\
T_\downarrow = \frac{|t_{1-}|^2}{k_\text{in}} \left[ k_{1-}(k_\text{in}) +
k_\text{SO} \langle \sigma_z \rangle_{1-} \right]\,,\label{prob:down}
\end{eqnarray}
\endnumparts
%
%
where $k_\text{in}$ is the injection momentum and
$k_{1+}(k_\text{in})$ and $k_{1-}(k_\text{in})$ are the momentum
relative to $k_\text{in}$ for the two spin resolved sub-bands. The
factors $ \langle \sigma_z \rangle_{1+}$ and $ \langle \sigma_z
\rangle_{1-}$ are the expectation values of $\sigma_z$ on the two spin
resolved sub-band wave functions and are defined as
%
%
\begin{equation}
\langle \sigma_z \rangle_{1\pm} = \pm \frac{1-f_1(\pm k_{1\pm})^2 }{1+
f_1(\pm k_{1\pm})^2} 
\end{equation}
%
%
where the function $f_1$ has been provided with
Eq.~(\ref{fun:f:one}). Because of  the absence of SOI for $x<0$ the
reflection probabilities are simply defined as $R_\uparrow =
|r_{1+}|^2$ and $R_\downarrow = |r_{1-}|^2$. The system Hamiltonian
$\mathcal{H}_\text{hyb}$ is invariant under time-reversal symmetry
and, as consequence, the following relations hold: $R_\uparrow =
R_\downarrow$ and $T_\uparrow= T_\downarrow$.  This means that spin-up
and spin-down carriers are transmitted through the interface in the
same way.

We consider the injection of a spin-unpolarized mixture of carriers
with injection momentum $k_\text{in}$. In terms of density matrix we
have
%
%
\begin{equation}\label{rho:in}
\rho_\text{in} = \frac{1}{2} | \uparrow \,\rangle \langle \,\uparrow
|+\frac{1}{2} | \downarrow \,\rangle \langle \,\downarrow | 
\end{equation}
%
%
with the property that $\langle \sigma_z \rangle_\text{in}=
\text{Tr}\{\rho_\text{in} \sigma_z\}=0$. The density matrix of the
transmitted carriers is expressed by the relation
%
%
\begin{eqnarray}
\rho_\text{out} &= & \frac{T_\uparrow}{T_\uparrow+T_\downarrow} |1+
\rangle \langle 1+ | + \frac{T_\downarrow}{T_\uparrow+T_\downarrow}
|1- \rangle \langle 1- | \nonumber \\ 
&=& \frac{1}{2} |1+ \rangle \langle 1+ | + \frac{1}{2} |1- \rangle
\langle 1- | \label{rho:out}
\end{eqnarray}
%
%
where $| 1\pm\rangle$ are the wave functions of the spin resolved
sub-bands. We can now evaluate the polarization of the output
carriers, this is expressed by
%
%
\begin{eqnarray}
\langle \sigma_z \rangle_\text{out} &=& \frac{1}{2}  \langle \sigma_z
\rangle_{1+}  + \frac{1}{2} \langle \sigma_z\rangle_{1-} \nonumber \\ 
&=& \frac{1}{2} \left( \frac{1-f_1(k_{1+})^2 }{1+
f_1(k_{1+})^2}-\frac{1-f_1(- k_{1-})^2 }{1+ f_1(-k_{1-})^2} \right)
\label{polarization:one}.
\end{eqnarray}
%
%
%
%
\begin{figure}
        \centering 
        \includegraphics[scale=0.3]{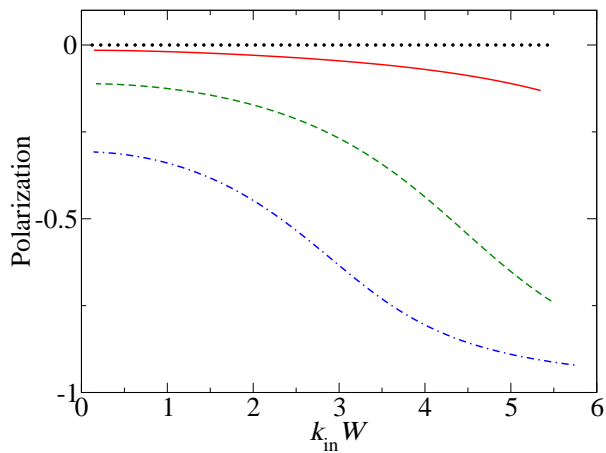}
        \caption{\label{pol:one}Polarization (\ref{polarization:one})
        as a function of the injection energy for
        $\bar{k}_\text{SO}=0$ (dotted line), $\bar{k}_\text{SO}=1$
        (solid line), $\bar{k}_\text{SO}=2$ (dashed line) and
        $\bar{k}_\text{SO}=3$ (dotted-dashed line).}
\end{figure}
%
%
Figure~(\ref{pol:one}) shows the polarization (\ref{polarization:one})
as a function of the injection energy for various values of the
SOI. The injection energy is limited within the first two
spin-resolved sub-bands. It is clear that when SOI is zero (dotted
line) there is no polarization, but as soon as the SOI is different
from zero, the polarization gets a finite value, which increases as a function
of SOI for a fixed energy. Those results are in accordance with
Governale and Z\" ulicke~\cite{governale:2002}, who showed a negative
polarization for carriers in the first two spin-resolved sub-bands and
positive injection energies.

\subsection{Wire with finite SOI region}
We consider a QW composed of three parts: two external
regions ($x<0$ and $x>L$) without SOI, and a central one ($0<x<L$)
where SOI is present. We consider the hybrid system Hamiltonian of
Eq. (\ref{eqhyb}), with $k_\text{SO}(x)=k_\text{SO} \theta(x)
\theta(L-x)$ and two $\delta$-like potentials of $x=0$ and $x=L$.  Also
in this case, for simplicity, the mass is assumed constant along the
wire. As for the single interface case, the spinor eigenstates of
$H_\text{hyb}$ are continuous at the interfaces, whereas their
derivatives have discontinuities in $x=0$ and $x=L$:
%
%
\numparts
\begin{eqnarray}
\psi(0^+)=\psi(0^-)\,,\label{eqhyb3a}\\
\left.\frac{\partial \psi(x)}{\partial x}\right|_{x=0^+} - 
\left.\frac{\partial \psi(x)}{\partial x}\right|_{x=0^-}=
(u-\ii\sigma_z k_\text{SO}) \psi(0)\,,\label{eqhyb3b} \\ 
\psi(L^+)=\psi(L^-)\,,\label{eqhyb3c}\\ 
\left.\frac{\partial \psi(x)}{\partial
x}\right|_{x=L^+}-\left.\frac{\partial \psi(x)}{\partial
x}\right|_{x=L^-}=  (u+\ii \sigma_z k_\text{SO}) \psi(L).
\label{eqhyb3d}
\end{eqnarray}
\endnumparts
%
%
As first step, we consider carriers with injection energy within the
first two spin-resolved sub-bands and with evanescent waves for the
following two. The Eqs. (\ref{eqhyb3a}-\ref{eqhyb3d}) reduce to a
set of two decoupled system of equations, with relevant terms
$t_{\text{R}1+}$, $t_{\text{R}1-}$, $r_{\text{L}1+}$ and
$r_{\text{L}1-}$. Those are practical for evaluating the transmission
and the reflection probabilities for spin-up and spin-down carriers.
As expected, because of the absence of SOI in the external regions we get
%
%
\numparts
\begin{eqnarray}
T_{\uparrow} = | t_{\text{R}1+}|^2\,\,\,,\,\,\, T_\downarrow =
|t_{\text{R}1-}|^2\,,\\ 
R_{\uparrow} = | r_{\text{L}1+}|^2\,\,\,,\,\,\, R_\downarrow =
|r_{\text{L}1-}|^2. 
\end{eqnarray}
\endnumparts
%
%
Due to time-reversal symmetry, it results that the value of the
transmission probability $T_\uparrow$ for spin-up incoming carriers is
equal to $T_\downarrow$ for incoming spin-down carriers.
It is relevant to study transport properties when an unpolarized
mixture of spin-up and spin-down carriers is injected into the
system. The incoming and the outgoing density matrices are described
by the expressions (\ref{rho:in}) and (\ref{rho:out}) respectively,
where now the states $|1\pm\rangle$ are the output wave functions in
the second region without SOI. As in the previous section we can
evaluate the polarization as the average value of the $\sigma_z$
operator and obtain as result
%
%
\begin{equation}
\langle \sigma_z \rangle _\text{out} = T_\uparrow -T_\downarrow = 0.
\end{equation}
%
%
The effect of spin polarization due to the first interface is
completely cancelled by the second one, therefore it is not possible to
observe any spin polarization~\cite{governale:2002}. This result can
be also derived by a symmetry consideration: let us consider the
scattering matrix $\mathcal{S}_\text{QW}$ of the QW. In absence of a
magnetic field time-reversal symmetry is preserved, therefore, as a consequence,
for $\mathcal{S}_\text{QW}$ holds the following important relation:
%
%
\begin{equation}\label{trs}
\mathcal{S}_\text{QW} = \Sigma_y \mathcal{S}_\text{QW}^\dag\Sigma_y
\end{equation}
%
%
where $\Sigma_y = \left( \begin{array}{c c}\sigma_y & 0_2 \\ 0_2 &
\sigma_y\end{array} \right)$. It is clear for Eq.~(\ref{trs}) that in
the case of only one conducting channel the spin-flip transmission
terms must be zero and as consequence the polarization is
absent~\cite{zhai:2005}.

As a further step, we study the transport properties when carriers are
injected also within the second two spin-resolved sub-bands. In this
limit all the evanescent modes are transformed in conducting
ones. Using Eqs. (\ref{rho:in}) and (\ref{rho:out}) for the incoming
and the out-coming density matrix, the polarization is expressed by
%
%
\begin{eqnarray}
\langle \sigma_z \rangle_\text{out}  & = & \frac{1}{2} \left(
\frac{T_{1+}(0)+T_{2+}(0)-T_{1-}(0)-T_{2-}(0)}{T_{1+}(0)+T_{2+}(0)+T_{1-}(0)+
T_{2-}(0)}+ \right. \nonumber \\ 
& & +
\left. \frac{T_{1+}(\pi/2)+T_{2+}(\pi/2)-T_{1-}(\pi/2)-
T_{2-}(\pi/2)}{T_{1+}(\pi/2)+T_{2+}(\pi/2)+T_{1-}(\pi/2)+T_{2-}(\pi/2)}
\right),\label{for:poltot}
\end{eqnarray}
%
%
where ``$0$'' and ``$\pi/2$'' denote incoming up- and down-carriers
respectively.
%
%
\begin{figure}[t]
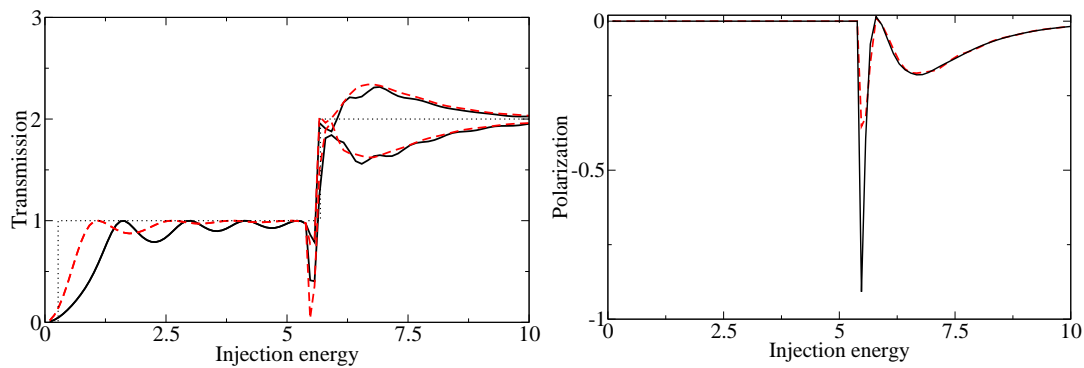

\begin{minipage}[c]{0.45\textwidth}
   \centering
   \includegraphics[width=\textwidth]{fig5a} 
\end{minipage}
\begin{minipage}[c]{0.45\textwidth}
   \centering
   \includegraphics[width=\textwidth]{fig5b} 
\end{minipage}
   \caption{(Left Panel) Spin resolved transmissions as a function of
   the injection energy and for two different values of the
   transparency of the barriers: $u=1.0$ (solid lines) and $u=0.1$
   (dashed lines).  The dotted line indicates the opening 
   of the sub-bands in absence of SOI.   The dimensionless SOI is $\bar{k}_\text{SO}=1.4$ and
   the central region is $L=3W$. (Right Panel) Spin polarization
   $\langle \sigma_z \rangle$ as a function of the injection energy
   and for two different value of the $\delta$-barrier
   transparency. The same parameters as in the left panel are used.}
   \label{pol_ana}
\end{figure}
%
%

In Fig.~\ref{pol_ana} (left panel) we show the spin resolved
transmissions as a function of the injection energy below and above
the bottom of the second two sub-bands and for two values of the
transparency $u$ of the interfaces (strength of the $\delta$-function potential 
in Eq.~\ref{eqhyb}).  The behavior below the threshold
clearly shows Fabry-Perot oscillations due to multiple reflection
effects within the central region, moreover the strength of those
oscillations is decreasing for increasing transparency of the two
$\delta$-barriers. A second important feature of the behavior is that
up to the threshold, owing to the time-reversal symmetry, the two spin
resolved components possess the same value. When the injection energy
is crossing the bottom of the second two sub-bands, a new behavior 
appears. The two spin resolved
transmissions take different values, the difference between those two
is bigger near to the sub-band threshold and is decreasing with
increasing energy. This can be well understood if we relate this
phenomenon to the sub-band hybridization. As shown in
Figs.~\ref{fig_1_a} and \ref{fig_2_a} the SOI strongly modifies
the parabolic-like behavior of the spin resolved sub-bands when new
sub-bands are opening and this effect is vanishing at higher
energies. In contrast to the results of Governale and Z\"
ulicke~\cite{governale:2002}, Fig.~\ref{pol_ana} clearly shows how
polarization effects can manifest themselves when also the second two
spin-resolved sub-bands are involved into the transmission
mechanism. This is more clear in Fig.~\ref{pol_ana} (right panel),
where the polarization as a function of the injection energy is
shown. The polarization is zero below the bottom of the second two
sub-bands and has oscillating behavior above it.  This particular
shape of the polarization can be understood also in terms of formation
of spin-dependent bound states at energies closer to the opening of
the second two sub-bands. Those bound-state oscillations are clearly
visible in the polarization pattern and it is evident how they are
enhanced when the $\delta$-barrier transparency is decreased.

The effects on spin transport due to higher sub-bands can be
numerically verified. For this purpose we employ a tight-binding model
of the Hamiltonian (\ref{eq1}), the spin-dependent scattering
coefficients are obtained by projecting the corresponding
spin-dependent Green function of the open system onto an appropriate
set of asymptotic spinors defining incoming and outgoing channels.  A
real-space discretization of the Schr\"odinger equation in combination
with a recursive algorithm for the computation of the corresponding
Green function has been implemented~\cite{ferry:1997}. This formalism
allows a convenient treatment of different geometries as well as
different sources of scattering within the same
framework~\cite{frustaglia:2004}.

%
%
\begin{figure}
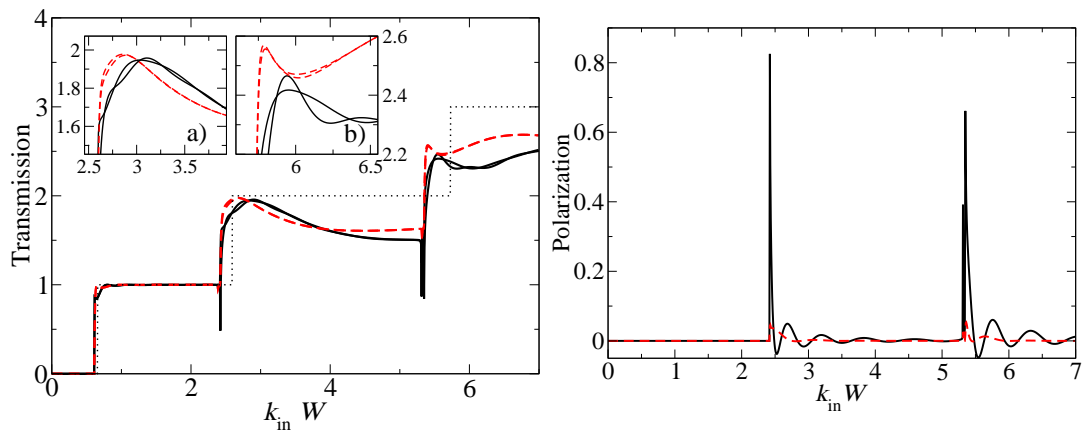
 
\begin{minipage}[c]{0.45\textwidth}
   \centering
   \includegraphics[width=\textwidth]{fig6a} 
\end{minipage}
\begin{minipage}[c]{0.45\textwidth}
   \centering
   \includegraphics[width=\textwidth]{fig6b} 
\end{minipage}
   \caption{(Left Panel) Numerical spin resolved transmissions as a
   function of the injection energy and for two different values of
   the switching region of SOI: $L_\text{sr}W^{-1}\sim0.07$ (solid lines) and
   $L_{\text{sr}}W^{-1}\sim0.7$ (dashed lines). The dotted line indicates the opening 
   of the sub-bands in absence of SOI.  The dimensionless SOI is
   $\bar{k}_\text{SO}=1.4$ and the central region is
   $L=3W+2L_{\text{sr}}$. In the insets a) and b) are shown the magnification
    for the steps relative to the opening of the second and the third sub-band, respectively. 
    (Right Panel) Spin polarization $ \langle
   \sigma_z \rangle$ as a function of the injection of energy and for
   the two different values of $L_\text{sr}W^{-1}$. The same parameters
   as in the left panel are used.}  \label{pol_num}
\end{figure}
%
%
The $\delta$-barriers are introduced in the tight-binding calculation
dividing the central region in three parts in order to modulate the
switching region $L_\text{sr}$ for the SOI. A long switching region
will correspond to a high transparency, and \emph{vice versa}.  In
Fig.~\ref{pol_num} (left panel) we show the spin-resolved transmission
as a function of the injection energy and for two different lengths of
the switching region. The system parameters are chosen in order to
have three active sub-bands in the case of injection within the
highest energy allowed by the tight-binding approximation of the
Hamiltonian (\ref{eq1}).  Injecting carriers within the first two
spin-resolved sub-bands reproduces the result coming from time-reversal symmetry: 
spin-up and spin-down transmissions coincide. As soon as
the second two sub-bands are opened we can observe a difference in
their values. This difference tends to decrease for increasing energy but is, then,
enhanced by the opening of the third two
sub-bands~\cite{sanchez:2006:note}. This is clear from 
Fig.~\ref{pol_num} (right panel), where we show the polarization as
a function of the injection energy. Here it is also possible to
observe how the strength of the polarization oscillations in strongly
reduced changing the length of the switching region.

\begin{figure}
\centering
\includegraphics[width=0.6\textwidth]{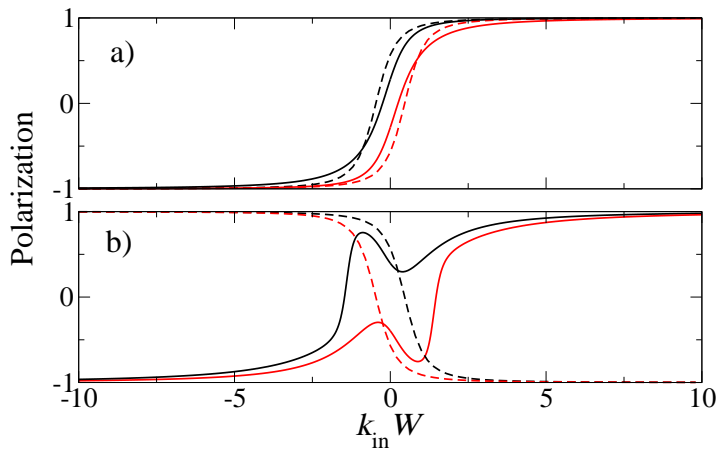}
\caption{\label{comparason:pol}Polarization as a function of the
injection energy for the two-band (dashed line) and the $N$-band
(solid line) models, respectively. Panel a) first sub-band, Panel b) second
sub-band. For both the Panels $N=50$.}
\end{figure}

The polarizations at the opening of the second two spin-resolved
sub-bands in the case of the two-band model (Fig.~\ref{pol_ana}) and
for the numerical tight-binding method (Fig.~\ref{pol_num}) show an
opposite value. 
This can be well understood if we consider the polarization
within the first and the second sub-band for an infinite long wire in
the two-band and in the $N$-band models, respectively. In
Fig.~\ref{comparason:pol} we show the polarization for the first
sub-band (upper Panel), it is evident that there is a small deviation
of the two-band model (dashed lines) from the $N$-band model. But this
deviation is stronger for the second sub-band (lower Panel) resulting
in completely opposite value at large energies.
This well reveals how the two-band model can only
qualitatively reproduce the exact calculation, but is failing in the
quantitative estimation. 
%
%
%
%
\section{Conclusions}
We have studied the properties of a QW in the presence of Rashba
SOI. The spectral properties have been investigated both from the
point of view of the exact wave functions and within the
second-order perturbation theory approach. Furthermore, a numerical
diagonalization procedure has been implemented in order to study the
spectral properties for the system in presence of $N$ sub-bands. We
have used this last method with two sub-bands within the so called
two-band model. We have shown that this is a good
description for the full properties of the first sub-band but not of
the second one. This model has also been used in order to study the
transport properties of a system with an interface between a region
with and without SOI. We have shown that the interface is spin
selective and that the crossing of unpolarized carriers through the
interface can give rise to a non-zero polarization. We have studied
spin transport also injecting carriers within the first and the second
sub-band. We have shown that at the opening of each new channel the
sub-band hybridization can give rise to spin selective bound states
reflecting in a oscillating spin polarization. Those results have been
confirmed by numerical calculation obtained within the tight-binding
approximation.

\ack
We acknowledge \.I. Adagideli, M. Scheid and M. Strehl for useful
discussions. D.B. wants to thank the hospitality at the
Max-Planck-Institut f\" ur Physik komplexer Systeme (Dresden) where
this paper was partially written and the financial support of Deutsche
Forschungsgemeinschaft within the cooperative research center SFB 689
``Spin phenomena in low dimensions''.


\Bibliography{99}
\bibitem{zutic:2004} I. \v Zut\'\i c, J. Fabian, and S. Das Sarma,
Rev. Mod. Phys. {\bf 76}, 323 (2004).

\bibitem{wolf:2001} S.A. Wolf \textit{et al.}, Science \textbf{294},
1488 (2001).

\bibitem{Datta:1990}  S. Datta and B. Das, Appl. Phys. Lett. {\bf 56},
665 (1990). 

\bibitem{schmidt:2000} G. Schmidt \textit{et al.}, Phys. Rev. B
\textbf{62}, 2790 (2000).

\bibitem{jedema:2001} F.J. Jedema, A.T. Fillip, and B.J. van Wess,
Nature \textbf{410}, 345 (2001).

\bibitem{valenzuela:2006} S.O. Valenzuela and M. Tinkham, Nature \textbf{442},
176 (2006).

\bibitem{dresselhaus:1955} G.~Dresselhaus, Phys. Rev. \textbf{100}, 580 (1955).

\bibitem{rashba:1960} E. Rashba, Fiz. Tverd. Tela (Leningrad)
\textbf{2}, 1224 (1960), [Sov. Phys. Solid State 2, 1109 (1960)].

\bibitem{nitta:1997} J. Nitta, T. Akazaki,
H. Takayanagi, and T. Enoki, Phys. Rev. Lett. \textbf{78}, 1335 (1997).

\bibitem{schaepers:1998} Th. Sch\"
apers, J. Engels, T. Klocke, M. Hollfelder, and H. L\" uth,
J. Appl. Phys. \textbf{83}, 4324 (1998). 

\bibitem{grundler:2000} D. Grundler, Phys. Rev. Lett. \textbf{84}, 
6074 (2000).

\bibitem{schaepers:2004} Th. Sch\" apers, J. Knobbe, and V. A. Guzenko, 
Phys. Rev. B \textbf{69}, 235323 (2004)

\bibitem{Mireles:2001}  F.Mireles and G.Kirczenow, Phys.Rev.B {\bf 64, }024426
(2001).

\bibitem{governale:2002}  M. Governale, U. Z\"{u}licke, Phys. Rev. B {\bf 66},
073311 (2002).

\bibitem{zhai:2005} Feng Zhai, and H. Q. Xu,
Phys. Rev. Lett. \textbf{94}, 246601 (2005). 

\bibitem{Marigliano1}  V. M. Ramaglia, D. Bercioux, V. Cataudella, G.De
Filippis, C.A. Perroni and F. Ventriglia, Eur. Phys. J.  B {\bf 36, }365 (2003).

\bibitem{marigliano:2004}  V.M. Ramaglia, D. Bercioux, V. Cataudella, G.De
Filippis, and C.A. Perroni, J. Phys.: Condens Matter \textbf{16}, 9143 (2004).

\bibitem{bercioux:2004} D. Bercioux, and V. M. Ramaglia, Superlattices Microstruct. \textbf{37}, 337 (2005).

\bibitem{Laurincikas:2003} A. Laurincikas and Ramunas Garunkstis
\emph{The Lerch Zeta-Function} (Kluwer Academic Pub). 

\bibitem{ferry:1997} D.~K. Ferry and S.~M. Goodnick, {\sl Transport in
Nanostructures} (Cambridge University Press, 1997).

\bibitem{frustaglia:2004} D. Frustaglia, M. Hentschel, and K. Richter,
Phys. Rev. B \textbf{69}, 155327 (2004).

\bibitem{sanchez:2006:note} Bound states owing to Fano resonances  in quantum wire with Rashba spin-orbit interaction have been recently 
investigated also by S\' anchez and Serra, Phys. Rev. B \textbf{74}, 153313 (2006).

\end{thebibliography}
\end{document}